\documentclass[fleqn,10pt]{wlscirep}
\usepackage[utf8]{inputenc}
\usepackage[T1]{fontenc}

\title{Single-Qubit Reaped Quantum State Tomography}

\author[1,*]{Mahn-Soo Choi}

\affil[1]{Department of Physics, Korea University, Seoul 02841, Republic of Korea}

\affil[*]{choims@korea.ac.kr}


\begin{abstract}
Quantum state tomography is the experimental procedure of determining an
unknown state. It is not only essential for the verification of resources and
processors of quantum information but is also important in its own right with
regard to the foundation of quantum mechanics.
Standard methods have been elusive for large systems because of the enormous
number of observables to be measured and the exponential complexity of
data post-processing.
Here, we propose a new scheme of quantum state tomography that requires the measurement of only three observables (acting jointly on the system and  pointer) regardless of the size of the system. The
system is coupled to a ``pointer'' of single qubit, and the wavefunction of
the system is ``reaped'' onto the pointer upon the measurement of the system.
Subsequently, standard two-state tomography on the pointer and classical
post-processing are used to reconstruct the quantum state of the system.
We also developed an efficient and scalable iterative maximum likelihood
algorithm to estimate states from statistically incomplete data.
\end{abstract}

\begin{document}

\newcommand\gray{\color{gray}}
\newcommand\red{\color{red}}
\newcommand\set[1]{{\textstyle\{#1\}}}
\newcommand\bra[1]{\mathinner{\langle{\textstyle#1}\rvert}}
\newcommand\ket[1]{\mathinner{\lvert{\textstyle#1}\rangle}}
\newcommand\braket[1]{\mathinner{\langle{\textstyle#1}\rangle}}
\newcommand\BfC{\mathbb{C}}
\newcommand{\hatH}{\hat{H}}
\newcommand{\hatI}{\hat{I}}
\newcommand{\hatO}{\hat{O}}
\newcommand{\hatP}{\hat{P}}
\newcommand{\hatR}{\hat{R}}
\newcommand{\hatU}{\hat{U}}
\newcommand{\hatV}{\hat{V}}
\newcommand{\hatW}{\hat{W}}
\newcommand{\hatX}{\hat{X}}
\newcommand{\calH}{\mathcal{H}}
\newcommand{\calM}{\mathcal{M}}
\newcommand{\varL}{\mathcal{L}}

\flushbottom
\maketitle

\thispagestyle{empty}

\section*{Introduction}

To develop reliable quantum resources and devices for quantum
information processing, it is crucial to verify their actual performance.
This is achieved at various levels, such as quantum process
tomography \cite{Chuang97a} and quantum detector tomography,
\cite{Lundeen09a,Zhang12a} and at the most fundamental level is the quantum
state tomography, which is the procedure of experimentally determining an unknown quantum state.\cite{Vogel89a}
Quantum state tomography is of great interest in its own right with regards to
the foundation of quantum mechanics as well.

In the standard formulation,\cite{Paris04a} quantum state tomography is
accomplished by performing repeated measurements of numerous non-commuting
observables on many systems prepared in the same states. As a matter of
principle, if the set of non-commuting observables is complete and the
measurements are repeated infinitely many times, one can build up a
comprehensive description of the quantum state by post-processing the
measurement statistics.\cite{James01a,Altepeter04a} It is recapitulated by
the three requirements of the standard quantum state tomography: (i) a
complete set of observables to be measured (so-called ``quorum''), (ii) 
accurate measurement statistics, and (iii) efficient post-processing.
In practice, the requirement of measuring a complete set of observables causes
overwhelming experimental obstacles, which affects the other requirements. Technical reasons and other difficulties may prevent some observables from being measured experimentally.
For large systems, the number of required observables is
exponentially large and places a serious limit on the number of repetitions of
measurements (which is finite anyway in reality). Both issues lead to
incomplete measurement statistics and/or limited accuracy of measurement statistics. Furthermore, even if
reasonably accurate measurement statistics are attained, the complexity of
post-processing itself is exponentially high for large systems.
To overcome such difficulties in \emph{exact} quantum state tomography,
various statistical methods have been developed to estimate quantum states, such as the maximum likelihood estimation \cite{Hradil97a,Hradil04a} and
Bayesian estimation \cite{Fuchs04a,Blume-Kohout10a,Rau10a} methods.
Notably, most statistical estimation methods, including the ML and Bayesian
approaches, are highly nonlinear procedures and generally suffers from high
complexity for large systems.

Here, we propose a new quantum state tomography scheme that requires the measurement of only three \emph{observables} regardless of the system size.\cite{endnote:1} In this
scheme, the system is coupled to a ``pointer'' of a single two-level quantum
system (i.e., ``qubit''), and the wavefunction of the system is ``reaped''
onto the pointer upon the measurement of a single observable on the system.
The subsequent standard quantum state tomography on the pointer and classical
post-processing reconstruct the quantum state of the system, where the
classical post-processing requires matrix inversion.  We refer to this scheme as \emph{single-qubit reaped} (or \emph{pointer-reaped}) quantum state
tomography.
We have also developed an iterative maximum likelihood (ML) estimation algorithm that is adaptable to the single-qubit reaped scheme.
The iterative ML estimation algorithm is demonstrated by numerical simulations with several interesting quantum states, such as the GHZ, W, and Dicke states.
Furthermore, by matrix product state (MPS) representations, the
iterative ML algorithm is scalable and provides an efficient method to obtain
MPS estimates for the mixed states of large systems.
The MPS pure state estimate for
the mixed state determines the lower bound of the fidelity between the pure
and mixed states and can be used to experimentally verify the purity of
the laboratory-generated states.\cite{Cramer10a}

\section*{Results}

\subsection*{Exact Tomography}

Consider a system of $n$ particles, each of which has dimension $d$, such that
the total dimension of the system is
\begin{math}
N := d^n.
\end{math}
Let $\set{\ket{x}|x=0,\cdots,N-1}$ be the computational basis of the Hilbert space.
Suppose that we have an ensemble of such systems, identically prepared in the unknown state
\begin{math}
\ket\psi = \sum_{x=0}^{N-1}\ket{x}\psi_x
\end{math}
with the ``wavefunction'' $\psi_x\in\BfC$, where $\BfC$ is the set of complex
numbers. We assume that $\psi_0\neq 0$ without a loss of generality (a physical
state cannot be a null vector).  Our proposed scheme is illustrated in the two
equivalent quantum circuits in Fig.~\ref{TomographyPaper::fig:1}. We discuss these procedures in the following order:

\begin{figure}[b!]
\centering
\includegraphics[width=7cm]{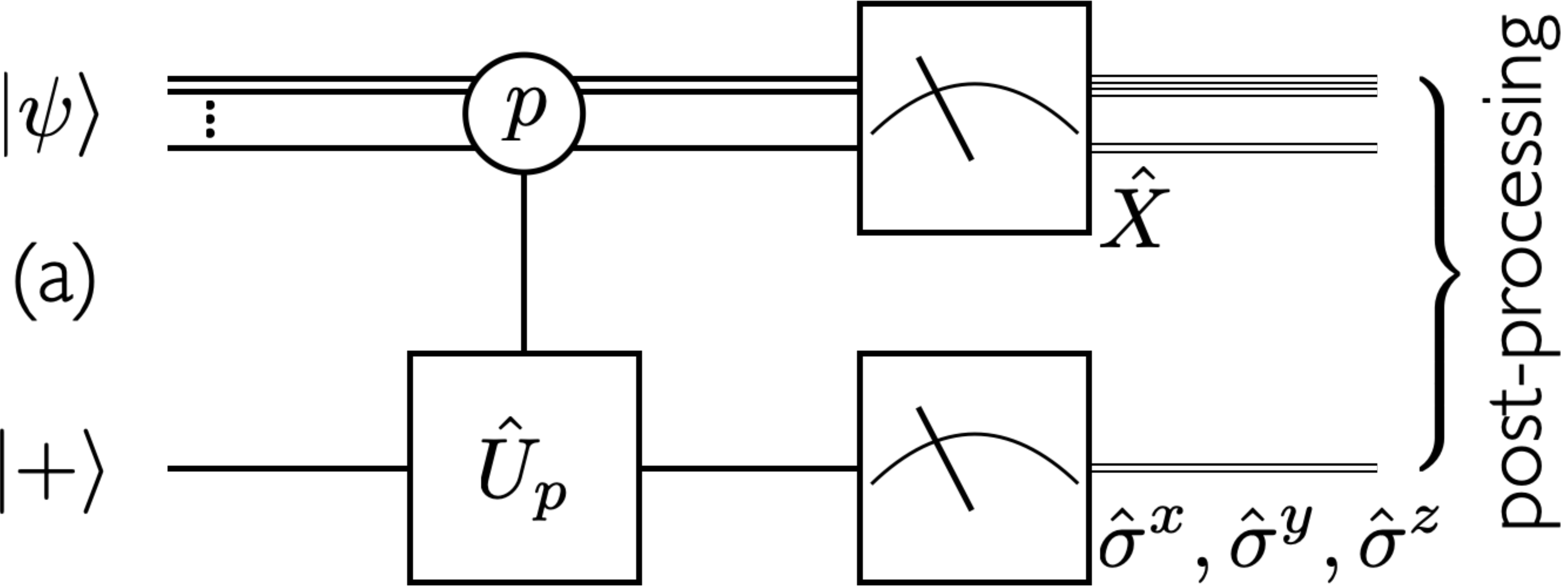}\quad
\includegraphics[width=7cm]{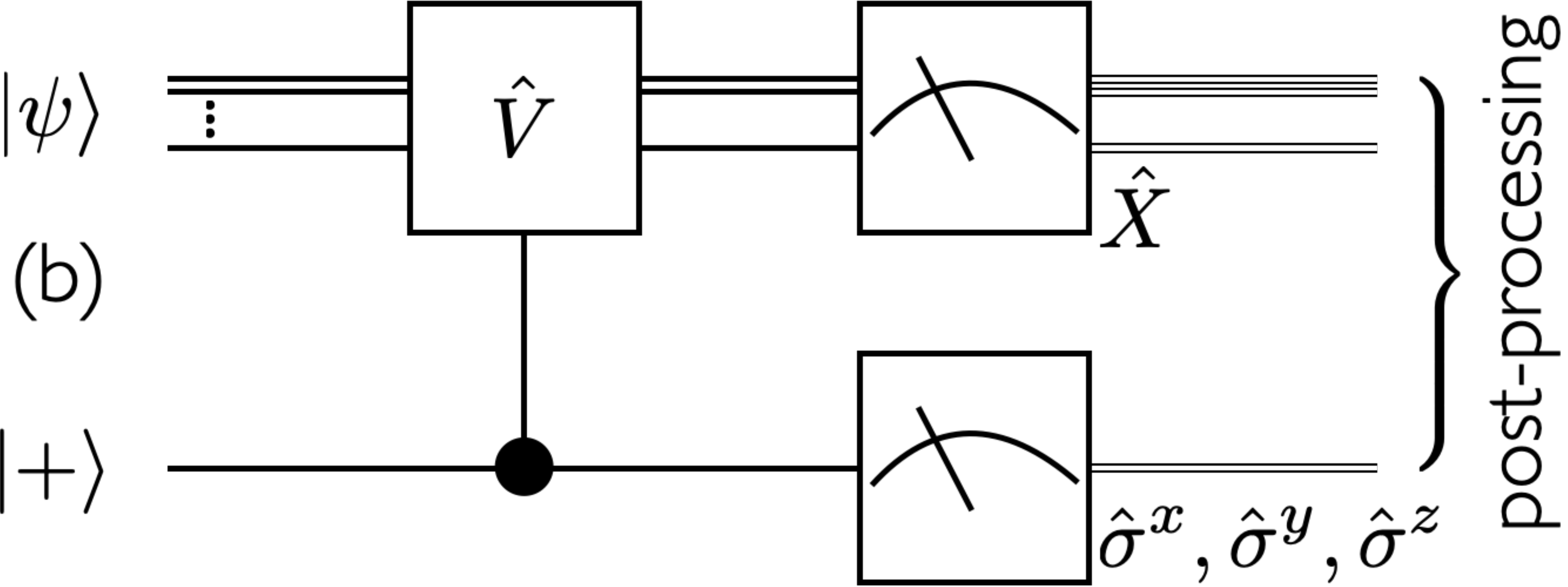}
\caption{Two equivalent schematics of the single-qubit reaped quantum state
  tomography. (a) The system-pointer interaction is described by the
  $p$-dependent conditional phase shift
  $\hatU_p:=\ket{0}\bra{0}+e^{ip\theta}\ket{1}\bra{1}$ on the pointer. (b) The
  system-pointer interaction is regarded as the pointer-controlled unitary
  operator $\hatV:=e^{i\theta\hatP}$ on the system.  The measurement on the
  system measures the eigenvalues $x$ of the observable
  $\hatX:=\sum_xx\ket{x}\bra{x}$ whereas the measurement on the pointer
  measures the eigenvalues of the Pauli operators $\hat\sigma^x$,
  $\hat\sigma^y$, or $\hat\sigma^z$.}
\label{TomographyPaper::fig:1}
\end{figure}

First, we select a qubit as the ``pointer''.
The pointer plays a central role in the proposed scheme.
Initially, we prepare the pointer in the state $\ket{+}:=(\ket{0}+\ket{1})/\sqrt{2}$, where
$\ket{0}$ and $\ket{1}$ are the computational basis states of the pointer such that
the initial state of the system plus pointer is given by
\begin{math}
\ket\Psi = \sum_x\ket{x}\psi_x\otimes\ket{+}.
\end{math}

Next, we couple the system and pointer for a certain time, which is
assumed to be sufficiently short compared to the typical time scales of the
system and pointer. This interaction can be described by a unitary
operator of the form~\cite{endnote:2}
\begin{math}
\hatU_\mathrm{int} = \exp\left(i\theta\hatP\otimes\ket{1}\bra{1}\right),
\end{math}
where $\hatP$ is an observable of the system.
For the sake of physical implementation in actual experiments, one can take
two different but equivalent views of $\hatU_\mathrm{int}$.
One can represent $\hatU_\mathrm{int}$ with the phase shift on the pointer
\emph{conditioned} on the system observable $\hatP$.  To observe this more
explicitly, let $\ket{p}$ be the eigenstate of the observable $\hatP$
belonging to the eigenvalue $p$ and rewrite $\hatU_\mathrm{int}$ as
\begin{math}
\hatU_\mathrm{int} = \sum_p\ket{p}\bra{p}\otimes\hatU_p
\end{math}
with the $p$-dependent phase shift
\begin{math}
\hatU_p := \ket{0}\bra{0} + e^{ip\theta}\ket{1}\bra{1}
\end{math}
on the pointer.
This interpretation is depicted in the quantum circuit
representation in Fig.~\ref{TomographyPaper::fig:1} (a) and is analogous to the
conventional von Neumann picture of the measurement of the observable
$\hatP$. One important difference is that the pointer here is only of two
dimensions and is insufficient to directly discriminate the $N$ eigenvalues, $p$, of $\hatP$.
On the other hand, noting that
\begin{math}
\hatU_\mathrm{int}=\hatI\otimes\ket{0}\bra{0}+\hatV\otimes\ket{1}\bra{1}
\end{math}
with $\hatI$ being the identity operator and $\hatV:=e^{i\theta\hatP}$, one can regard it as a pointer-controlled unitary operator $\hatV$ acting on the system.
This picture is illustrated in the quantum circuit in
Fig.~\ref{TomographyPaper::fig:1} (b) and is analogous to the quantum phase
estimation circuit for a unitary transformation ($\hatV$ in the present
case).\cite{Kitaev02a}
Throughout this paper, we will mainly consider the latter interpretation for
convenience.
After the unitary interaction, the total state becomes
\begin{equation}
\label{TomographyPaper::eq:1}
\hatU_\mathrm{int}\ket\Psi
= \sum_{xy}\ket{x}\otimes
\frac{\ket{0}\delta_{xy}\psi_y + \ket{1}V_{xy}\psi_y}{\sqrt{2}},
\end{equation}
where $V$ is the matrix representation of $\hatV$ in the computational basis,
\begin{equation}
V_{xy} := \braket{x|\hatV|y}
= \sum_{p}\braket{x|p}e^{ip\theta}\braket{p|y} \,.
\end{equation}

We then measure the eigenvalues of the observable
$\hatX:=\sum_xx\ket{x}\bra{x}$ in the system.
When the measurement outcome is $x$, the (unnormalized) pointer state is reduced to
\begin{equation}
\label{TomographyPaper::eq:4}
\ket{\phi_x} = \ket{0}\psi_x + \ket{1}\sum_{y}V_{xy}\psi_y \,.
\end{equation}
Equation~\eqref{TomographyPaper::eq:4} reveals the key idea of the proposed scheme: the wavefunction $\psi_x$ appears in the two expansion coefficients and can be determined by the standard quantum state tomography by measuring three
independent observables, that is, the Pauli operators $\hat\sigma^x$,
$\hat\sigma^y$, and $\hat\sigma^z$ in the pointer.
One tricky point is that naive two-state tomography does not fix the overall
phase, which is necessary to fix the relative phases of $\psi_x$ for different
values of $x$. We now provide a careful tomographic reconstruction procedure [see Eq.~\eqref{TomographyPaper::eq:7}] that is not hindered by this tricky issue.

Physically, the two-step procedure for the measurement of $\hatX$ on the system
and the subsequent quantum state tomography on the pointer is equivalent to
the measurement of the eigenvalues of three observables,
$\hatX\otimes\hat\sigma^z$, $\hatX\otimes\hat\sigma^x$, and
$\hatX\otimes\hat\sigma^y$.
For the purpose of mathematical analysis of measurement outcomes and maximum likelihood estimation process (see below),
it is convenient to describe the measurements
using the projective POVM elements
\begin{equation}
\label{TomographyPaper::eq:5}
\hat\Pi_{x,m} := \frac{1}{3}\hat\Pi_x\otimes\hat\Pi_m
\end{equation}
where $\hat\Pi_x=\ket{x}\bra{x}$, $\hat\Pi_m=\ket{m}\bra{m}$, and the
index $m\in\calM:=\set{0,1,+,-,L,R}$ refers to the eigenstates
$\ket{m}=\ket{0},\ket{1},\ket{+},\ket{-},\ket{L},\ket{R}$ of the Pauli
operators $\hat\sigma^z$, $\hat\sigma^x$, and $\hat\sigma^y$, respectively.
The joint probabilities
\begin{math}
P_{x,m}
=\bra{\Psi}\hatU_\mathrm{int}^\dag\hat\Pi_{x,m}\hatU_\mathrm{int}\ket{\Psi}
\end{math}
determine the ratio between the two coefficients,
\begin{equation}
\label{TomographyPaper::eq:6}
\frac{1}{\psi_x}\sum_{y=0}^{N-1}V_{xy}\psi_y
= \sqrt{\frac{P_{x,1}}{P_{x,0}}}e^{i\varphi_x} ,
\end{equation}
where
\begin{math}
\varphi_x := \arg[(P_{x,+}-P_{x,-}) + i(P_{x,L}-P_{x,R})] \,.
\end{math}
Owing to the normalization constraint, the $N$ relations in
Eq.~\eqref{TomographyPaper::eq:6} are not independent of each other.
Instead of directly imposing the normalization constraint,
one can just determine the ratio $\psi_x/\psi_0$.  This casts the
relation~\eqref{TomographyPaper::eq:6} to the following set of $(N-1)$ linear
equations
\begin{equation}
\label{TomographyPaper::eq:7}
\sum_{y=1}^{N-1}\left\{
\sqrt{P_{x,1}}e^{i\varphi_x}\delta_{xy} - \sqrt{P_{x,0}}V_{xy}
\right\}
\left(\frac{\psi_y}{\psi_0}\right)
= \sqrt{P_{x,0}}V_{x0}
\end{equation}
for $x=1,\cdots,N-1$. 
Given the experimentally determined measurement statistics $P_{x,m}$, solving
the linear equations yields the wavefunction $\psi_x$ (up to normalization).
There are several dangerous cases in which Eq.~\eqref{TomographyPaper::eq:7} cannot provide a unique solution. Avoiding or overcoming them is addressed in Methods.

One simple example is to select the local basis $\ket{x}$ such that
\begin{math}
\braket{x|p} = N^{-1/2}e^{2\pi{i} xk_p/N} \,,
\end{math}
where $k_p$ is the index of $p$ when the eigenvalues are arranged in an
ordered sequence. The computational basis $\ket{x}$ and the eigenstates $\ket{p}$ of
$\hatP$ are related by the quantum Fourier transform.\cite{Nielsen11a}
For a system consisting of qubits ($d=2$), another valuable example is the
system operator of the form
\begin{math}
\hatP = \sum_{j=1}^n\hat\tau_j^x \,,
\end{math}
where $\hat\tau_j^x:=(\ket{0}\bra{1}+\ket{1}\bra{0})_j$ denotes the Pauli operator acting on the $j$th qubit.
This leads to a pointer-controlled unitary operator
\begin{equation}
\label{TomographyPaper::eq:2}
\hatV = e^{i\theta\hatP} =
\begin{bmatrix}
\cos\theta & i\sin\theta \\
i\sin\theta & \cos\theta
\end{bmatrix}^{\otimes n}
\quad (0<\theta<\pi/2)
\end{equation}
In this case, $\ket{x}$ and $\ket{p}$ are related to each other via the local Hadamard gates,
\begin{equation}
\left[\braket{x|p}\right]_{x,p=0,1,\cdots,2^n-1}
= H^{\otimes n}
\end{equation}
with
\begin{equation}
H := \frac{1}{\sqrt{2}}
\begin{bmatrix}
1 & 1 \\
1 & -1
\end{bmatrix}.
\end{equation}

\subsection*{Maximum Likelihood Estimation Algorithm}

Above, we have shown that, as a matter of principle, the single-qubit reaped
scheme can successfully reconstruct quantum states. It assumes an idealistic
situation where the probability distribution $P_{x,m}$ corresponding to the
POVM elements $\hat\Pi_{x,m}$ can be inferred from measurements. It is
possible only when the measurements are repeated infinitely many times, apart
from other technical imperfections; finite repetitions give rise to
statistical errors in the inferred probabilities $P_{x,m}$.  Obviously, the
statistical errors become more severe as the system size $n$ increases;
recall the number $6d^n$ of possible measurement outcomes $(x,m)$.  A popular
method to overcome such an issue is to follow the maximum
likelihood (ML) principle and seek the state that is most ``likely'' given
the experimental observations rather than the actual (and impossible-to-infer)
wavefunction.\cite{Hradil97a,Hradil04a,Altepeter04a,Rehacek07a,James01a} In
this section, we develop an iterative ML algorithm that can be combined with
the single-qubit reaping scheme discussed above.
We note controversies about the physically proper
estimation of quantum states from the experimental data,\cite{Rehacek07a,Blume-Kohout10a} and it would be valuable to develop other
statistical methods, such as Bayesian approaches, that are adaptable to the present tomography scheme.

Consider an ensemble of $F$ systems. Let $F_{x,m}$ be the number of
experimental observations corresponding to the POVM element $\hat\Pi_{x,m}$,
such that
\begin{math}
F = \sum_{x,m}F_{x,m}.
\end{math}
The ideal situation corresponds to the limit $F\to\infty$, where the relative
frequency $F_{x,m}/F$ gives the true probability $P_{x,m}$. For finite
size ($F<\infty$), $F_{x,m}/F$ only estimates $P_{x,m}$ approximately. The
observation statistics are governed by a multinomial distribution
\begin{equation}
\varL
= F!\prod_{x}\prod_{m\in\calM}\frac{(P_{x,m})^{F_{x,m}}}{F_{x,m}!} \,,
\end{equation}
where
\begin{equation}
P_{x,m}
=\bra{\Psi}\hatU_\mathrm{int}^\dag\hat\Pi_{x,m}\hatU_\mathrm{int}\ket{\Psi}
\end{equation}
is the probability of obtaining the result $(x,m)$ on the condition that the system plus pointer is prepared in the state
\begin{math}
\ket\Psi=\ket\psi\otimes\ket{+}.
\end{math}
We use the multinomial distribution $\varL$ as the likelihood function.
Generally, the likelihood function should depend on the speciﬁc measurement
apparatus and other experimental conditions. 
Here, we focus on the generic effects on statistical error, putting aside
specific technical issues.
The ML approach maximizes
\begin{equation}
\log\varL
= \sum_{x}\sum_{m\in\calM}F_{x,m}\log{P_{x,m}}
\end{equation}
(up to irrelevant terms) over all possible states $\ket\psi$ of the system
with the normalization constraint.
The wavefunction $\bar\psi_x$ that maximizes the likelihood function satisfies the extremal equation (see Methods for details)
\begin{equation}
\label{TomographyPaper::eq:8}
\sum_yW_{xy}\bar\psi_y = \bar\psi_x
\end{equation}
where the matrix $W$ is defined by
\begin{equation}
\label{TomographyPaper::eq:9}
W_{xy} := 
\braket{0|\hatR_x|0}\delta_{xz} +
\braket{0|\hatR_x|1}V_{xy} +
V^\dag_{xy}\braket{1|\hatR_y|0} +
\sum_zV^\dag_{xz}\braket{1|\hatR_z|1}V_{zy}
\end{equation}
and the $x$-dependent operator $\hatR_x$ on the pointer by
\begin{equation}
\label{TomographyPaper::eq:10}
\hatR_x := \sum_{m\in\calM} \frac{F_{x,m}}{P_{x,m}} \ket{m}\bra{m}.
\end{equation}
Formally, $\hatR_x$ is reminiscent of a similar operator (denoted by
$\hatR$) that appears in the iterative maximization algorithm adapted to the
standard tomography scheme.\cite{Hradil04a} In our case, $\hatR_x$ acts on
the pointer and not on the system itself.
In an ideal experiment where $F\to\infty$,
the true wavefunction indeed gives the
extremum solution, $\bar\psi_x=\psi_x$, as $\hatR_x=\hatI$.
In a realistic experiment with a finite-size ensemble ($F<\infty$), in general
$\bar\psi_x\neq\psi_x$, but $\bar\psi_x$ is simply the wavefunction most likely
for the given measurements data.

\begin{figure}[bt]
\centering
\includegraphics[height=25mm]{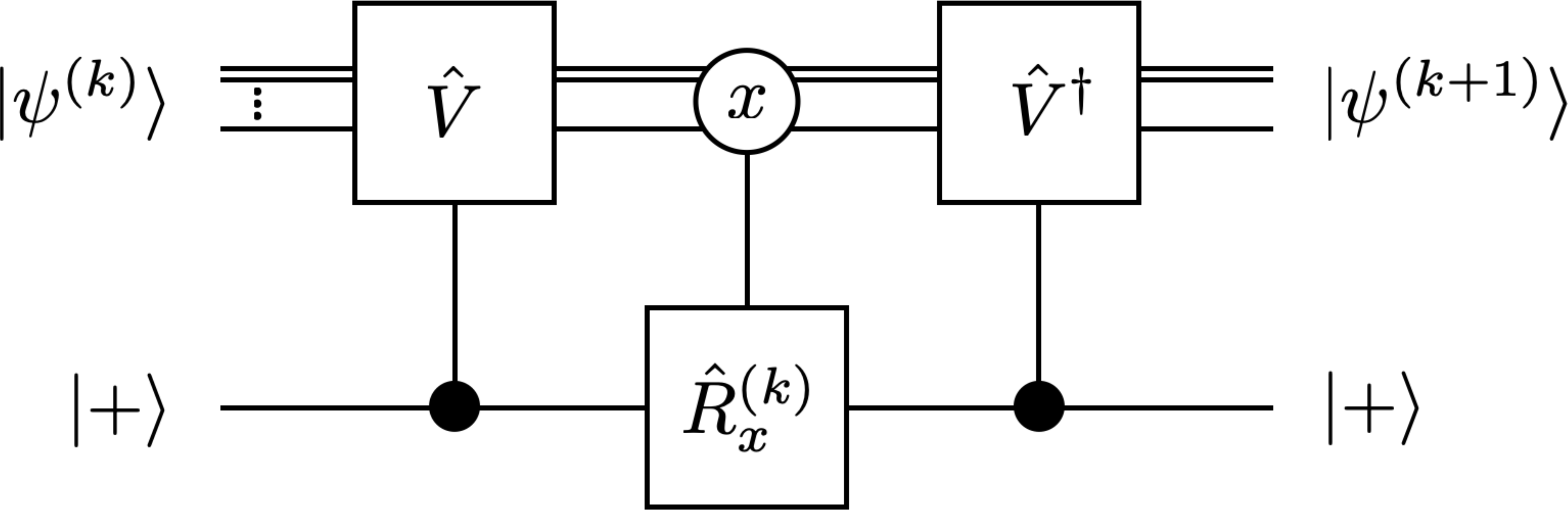}
\caption{Quantum circuit interpretation of the maximum likelihood
  iteration. The solid dot indicates the ``controlled''-$\hatV$ (or
  $\hatV^\dag$) acting only when the pointer is in the state $\ket{1}$ whereas
  the open circle indicates ``conditional''-$\hatR_x$ on the pointer
  conditioned on the state $\ket{x}$ of the system. Despite the quantum
  circuit interpretation, the iteration procedure is \emph{not} linear as the
  operator $\hatR_x[\psi^{(k)}]$ depends on the trial state
  $\ket{\psi^{(k)}}$.}
\label{TomographyPaper::fig:2}
\end{figure}

It should be noted that the operator $\hatR_x$ depends functionally on the state $\ket\psi$ through the probability $P_{x,m}$, and hence the extremum
equation~\eqref{TomographyPaper::eq:8} is \emph{nonlinear}.  Solving such a
nonlinear equation is unviable, particularly for large systems (involving a
large number of variables $\psi_x$).
Instead, we have developed an iterative algorithm
\cite{Rehacek07a,Hradil04a,Molina-Terriza04a,Lvovsky04a}.
First, we need to choose an initial trial wavefunction.  From the
pointer state $\ket{\phi_x}$ in Eq.~\eqref{TomographyPaper::eq:4} upon the
measurement readout $x$, it follows that the probability $P_{x,0}$ is directly
proportional to $|\psi_x|^2$. This implies that
$\ket{\psi^{(0)}} \propto \sum_x\ket{x}\sqrt{F_{x,0}/F}$ is a reasonable
choice.
At each iterative step $k$, the wavefunction $\ket{\psi^{(k)}}$ is updated using
the mapping
\begin{equation}
\label{TomographyPaper::eq:11}
\hatW[\psi^{(k)}]\ket{\psi^{(k)}} = \ket{\psi^{(k+1)}} \,,
\end{equation}
where the iteration generator $\hatW:=\sum_{xy}W_{xy}\ket{x}\bra{y}$ is constructed from the matrix $W$ in Eq.~\eqref{TomographyPaper::eq:9}.
Interestingly, the iteration procedure can be represented by the quantum
circuit shown in Fig.~\ref{TomographyPaper::fig:2},
which illustrates the crucial role of the pointer from another perspective.
The quantum circuit itself is not advantageous when one
evaluates the iterations directly. However, as we will observe later, it clearly reveals the simple mathematical structure of the iteration generator $\hatW$, which permits the scalability of the iterative algorithm.

\begin{figure}[bt]
\centering
\includegraphics[width=75mm]{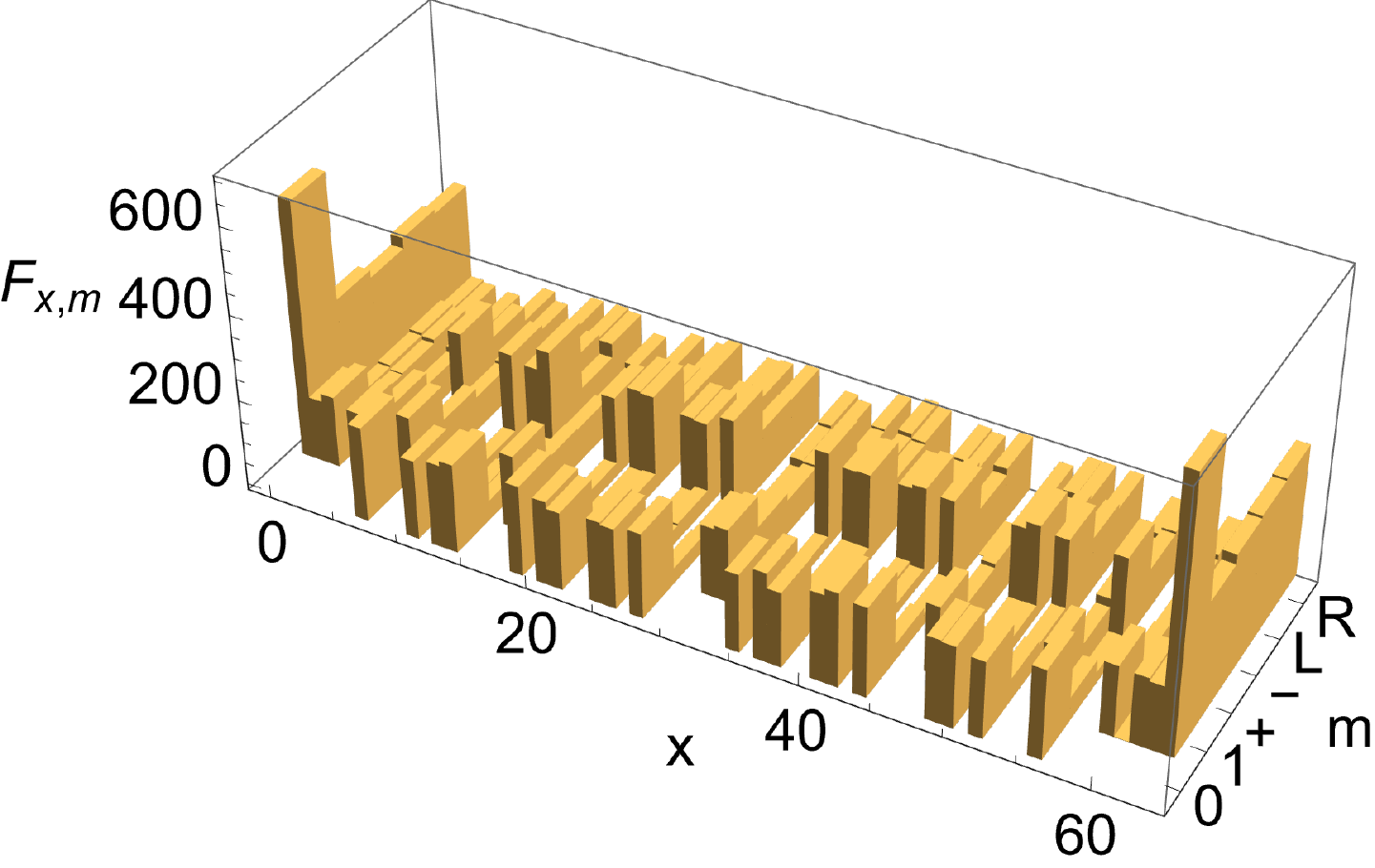} \quad 
\includegraphics[width=65mm]{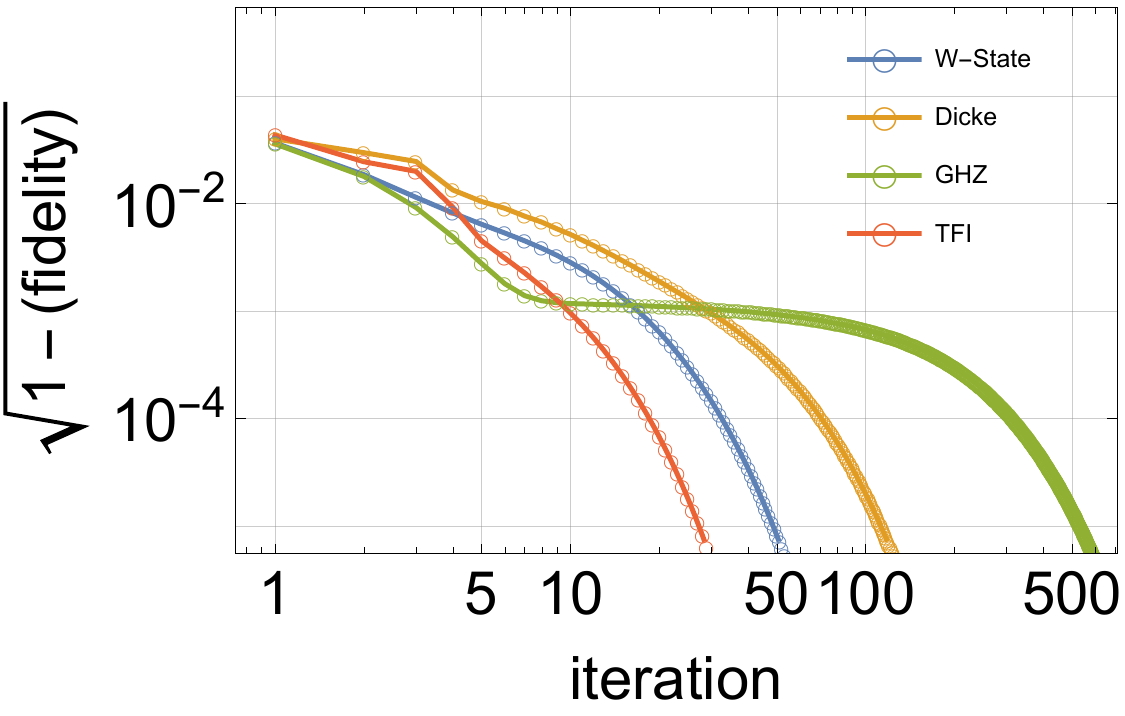} 
\caption{(a) Relative frequencies $F_{x,m}$ of the measurement readouts
  $(x,m)$ from the simulation with an ensemble of 24000 systems in the
  symmetric Dicke state with six qubits ($F=24000$, $n=6$, $d=2$).
  (b) Convergence behaviors of
  the iterative maximization procedure for different system states (the W
  state, Dicke state, GHZ state, and the ground state of the
  transverse field Ising model in the ordered phase)
  exhibited by the fidelity between the states
  from consecutive iterations. For all the four cases, the fidelities between
  the resulting states and the true wavefunctions, respectively, are better
  than 0.99.}
\label{TomographyPaper::fig:3}
\end{figure}

The convergence of no iterative ML algorithm has been analytically proven.\cite{Rehacek07a} However, in standard ML approaches,\cite{Baumgratz13b,Baumgratz13a} numerical tests have demonstrated
convergence for physically interesting states, and a diluted iterative
algorithm is available when the convergence is critical.\cite{Rehacek07a}
Here, we demonstrate the algorithm numerically using several examples for a system of six qubits ($n=6$ and $d=2$). The first example is the symmetric Dicke state
\begin{math}
\ket\psi = \sum'\ket{000111}/\sqrt{20},
\end{math}
where $\sum'$ refers to the summation over all permutations of the qubits.
We simulated the measurements for an ensemble of 24000 systems ($F=24000$)
  all prepared in the same state $\ket\psi$.
The resulting relative frequencies, $F_{x,m}$,
of the measurement readouts $(x,m)$
are shown in Fig.~\ref{TomographyPaper::fig:3} (a). We then obtained the ML
estimate $\ket{\psi^{(500)}}$ for the measurement data ($F_{x,m}$) through 500
iterations in accordance with \eqref{TomographyPaper::eq:11}.
As shown in Fig.~\ref{TomographyPaper::fig:3} (b, blue curve), the \emph{infidelity}
between the states from consecutive iterations was already less than
$10^{-5}$ after 150 iterations.
The fidelity,
\begin{math}
\left|\braket{\psi^{(200)}|\psi}\right|^2,
\end{math}
with the true wavefunction is larger then 0.997.

We performed similar simulations and made the ML estimates for the simulation results
for the W-state
\begin{math}
\ket\psi=(\ket{10 00 00}+\ket{01 00 00}+\cdots+\ket{00 00 01})/\sqrt{6},
\end{math}
the GHZ state
\begin{math}
\ket\psi=(\ket{000000}+\ket{111111})/\sqrt{2},
\end{math}
and the ground state of the transverse-field Ising model in the ordered phase.
Figure~\ref{TomographyPaper::fig:3} (b) corroborates the excellent
convergence for all those cases.
The fidelities between the ML
estimates and the respective true wavefunctions were also as good as 0.99 or
larger.

\subsection*{Scalability and Mixed States}

Each ML iteration in Eq.~\eqref{TomographyPaper::eq:11} involves
the multiplication of exponentially large matrices and vectors,
and the computational cost of many iterations
for the desired accuracy may still be high for large systems.
This can be overcome by means of matrix product state (MPS)
and matrix product operator (MPO) representations (see Methods).  We first examine the quantum
circuit shown in Fig.~\ref{TomographyPaper::fig:2} more closely to better understand the MPO structure of the iteration generator, $\hatW$.
Let $\hatW_\mathrm{tot}$ be the
extended operator acting on the system and pointer, which results in
\begin{math}
\hatW=\braket{+|\hatW_\mathrm{tot}|+}
\end{math}
when averaging over the pointer with the state $\ket{+}$.
$\hatW_\mathrm{tot}$ consists of the controlled-unitary
operator
\begin{math}
\hatI\otimes\ket{0}\bra{0}+\hatV\otimes\ket{1}\bra{1}
\end{math}
and the conditional-unitary operator
\begin{math}
\sum_x\ket{x}\bra{x}\otimes\hatR_x[\psi^{(k)}].
\end{math}
The former is an MPO with a bond dimension of 2
when the coupling observable $\hatP$ (and hence $\hatV$) is
local [Eq.~\eqref{TomographyPaper::eq:2} is an example].
The latter is also an
MPO with a finite bond dimension provided that the input state
$\ket{\psi^{(k)}}$ is an MPS with a finite bond dimension because an MPS only has finite correlations;\cite{Perez-Garcia07a,Schollwock11a}
see Methods. Therefore, $\hatW_\mathrm{tot}$, the product
of three MPOs, should be an MPO with a finite bond dimension, and so is $\hatW$
as it corresponds to a partial trace of an MPO.
Currently, the operation of an MPO
on an MPS can be efficiently evaluated.\cite{Perez-Garcia07a,Schollwock11a}
In summary, if the laboratory states are MPS, the iteration generator is
represented by an MPO, and the ML iterations in
Eq.~\eqref{TomographyPaper::eq:11} can be updated efficiently. Recently, a
formally similar iterative algorithm (from a different tomography scheme)
powered by MPO and MPS representations has been demonstrated in
detail.\cite{Baumgratz13a}

Because only a polynomial number of parameters is required for the MPS
representations, they span only a small portion of the entire Hilbert
space. However, it is well known that many states relevant to quantum
information processing, condensed matter physics, and other areas of physics
exist in the MPS form.  The ground states of the strongly correlated many-body
Hamiltonians as well as the cluster states are notable examples.

Moreover, as was pointed out recently,\cite{Cramer10a} the tomographic estimation of MPS pure states is valuable even when the system is in a mixed state. That is, it allows us to determine a lower bound on the fidelity
between the pure state estimate and mixed states compatible with the
experimental observations, thereby certifying the purity of the laboratory
state via experiments.
A scalable ML method has been proposed to directly reconstruct mixed states via local measurements,\cite{Baumgratz13b,Baumgratz13a} assuming that the
states are close to a MPS.
For their method, however, experimenters are required to measure many non-commuting observables whereas our scheme requires the measurement of only three observables
\begin{math}
\hatX\otimes\hat\sigma^x,
\end{math}
\begin{math}
\hatX\otimes\hat\sigma^y,
\end{math}
and
\begin{math}
\hatX\otimes\hat\sigma^z,
\end{math}
regardless of the system size.\cite{endnote:1}

\section*{Discussion}

A seemingly similar idea to couple the system with an ancillary system and
measure only one observable (over the entire system plus ancilla) has been previously proposed;\cite{Allahverdyan04a} this is the so-called ancilla-assisted quantum state tomography and has been demonstrated in recent experiments.\cite{Oren17a,Shukla13a}
However, their scheme required the ancilla to be as large
as or even larger than the system (one obvious advantage is that it
can directly estimate the density matrix of the system).
Moreover, no ML algorithm has been developed for their scheme.

The convergence of the ML iterations varies for different states.
For example, it is noted in Fig.~\ref{TomographyPaper::fig:3} (b) that the
convergence of the ML iterations is slower for the GHZ state (approximately
500 iterations are required for similar accuracy) than for other states.
Recalling the massive and long distance entanglement in the GHZ state, this
fact raises an interesting question about the relation between the convergence
behavior of our ML iterations and the properties (such as multi-partite
entanglement) of the state. We leave the relation as an inspiring open
question for future works.

\section*{Methods}

\subsection*{State-Reconstruction Equation}

Here, we derive the state-reconstruction equation~\eqref{TomographyPaper::eq:6}.
We begin with the (unnormalized) pointer state in Eq.~\eqref{TomographyPaper::eq:1}
\begin{equation}
\ket{\phi_x} = \ket{0}\alpha_x + \ket{1}\beta_x \,,
\end{equation}
where we have defined $\alpha_x:=\psi_x$ and $\beta_x:=\sum_yV_{xy}\psi_y$ for notational simplicity. We want to express the ratio $\beta_x/\alpha_x$ in terms of the joint probabilities $P_{x,m}$. The joint probabilities satisfy the following relationship:
\begin{subequations}
\begin{align}
P_{x,0} &= |\alpha_x|^2  ,\\
P_{x,1} &= |\beta_x|^2 ,\\
P_{x,+} - P_{x,-} &= \alpha_x^*\beta_x + \alpha_x\beta_x^* \,,\\
i(P_{x,L} - P_{x,R}) &= \alpha_x^*\beta_x - \alpha_x\beta_x^* \,.
\end{align}
\end{subequations}
Using the last two relations, one can obtain
\begin{equation}
P_{x,+}-P_{x,-} + i(P_{x,L}-P_{x,R}) = 2\alpha_x^*\beta_x \,.
\end{equation}
This implies that the relative phase between $\alpha_x$ and $\beta_x$, which is the essential part for quantum coherence effects, can be extracted by  combining the join probabilities on the left-hand side.
More explicitly, we express it as
\begin{equation}
\varphi_x := \arg\left[P_{x,+}-P_{x,-} + i(P_{x,L}-P_{x,R})\right] \,,
\end{equation}
and observe that
\begin{equation}
e^{i\varphi_x}
= \frac{ P_{x,+}-P_{x,-} + i(P_{x,L}-P_{x,R}) }{2|\alpha_x\beta_x|}
= \frac{\alpha_x^*}{|\alpha_x|}\frac{\beta_x}{|\beta_x|}
= \frac{\beta_x}{\alpha_x}
\frac{|\alpha_x|}{|\beta_x|}
= \frac{\beta_x}{\alpha_x}\sqrt{\frac{P_{x,0}}{P_{x,1}}},
\end{equation}
which is identical to Eq.~\eqref{TomographyPaper::eq:6}.
The physical implication of the above relation is that the probabilities $P_{x,0}$ and $P_{x,1}$ in the computational basis of the pointer give the relative \emph{magnitudes} of $\alpha_x$ and $\beta_x$, whereas the probabilities $P_{x,\pm}$ and $P_{x,L/R}$ give the relative \emph{phases} between them.

\subsection*{Dangerous Cases}

There are three dangerous cases where the wavefunction extraction scheme in
Eq.~\eqref{TomographyPaper::eq:8} may not give a unique solution:

(i) In the first case, $\hatP$ is compatible
with the computational basis, $\{\ket{x}\}$ ($[\hatX,\hatP]=0$). Then, $\ket{x}$ are essentially
eigenstates of $\hatP$, and the pointer state upon the measurement of $\hatX$
becomes
\begin{math}
\ket{\phi_x} = \psi_x(\ket{0}+\ket{1}e^{i\theta x}).
\end{math}
Because $\psi_x$ is an overall factor, it cannot be extracted.

(ii) In the second case, the unitary $\hatV$ is block diagonal (possibly after
simultaneous permutations of rows and columns) in a given basis. Suppose
that $\hatV=\hatV^{(1)}\oplus\hatV^{(2)}$ with $\hatV^{(1)}$ and $\hatV^{(2)}$
operating on orthogonal subspaces $\calH^{(1)}$ and $\calH^{(2)}$,
respectively, of $\calH^{(1)}\oplus\calH^{(2)}=\calH$.
Accordingly, any state $\ket\psi$ is decomposed into
\begin{math}
\ket\psi = \ket{\psi^{(1)}}\oplus\ket{\psi^{(2)}}.
\end{math}
Upon the measurement of $\hatX$, the pointer is cast to
\begin{equation}
\ket{\phi_x}
=\ket{0}\psi_x^{(\nu)}+\ket{1}\sum_y\hatV_{xy}^{(\nu)}\psi_x^{(\nu)}
\end{equation}
for $\ket{x}\in\calH^{(\nu)}$ ($\nu=1,2$). Therefore, in this case, one can assess $\psi_x^{(\nu)}/\psi_0^{(\nu)}$ by
applying the wavefunction extraction scheme
\eqref{TomographyPaper::eq:7}
for each sector $\nu$.
However, it is impossible to extract the phase relations between different sectors.

(iii) The third case is a special case where $\ket\psi$ happens to be an
eigenstate of $\hatP$ (i.e., $\hatV$) belonging to a \emph{degenerate}
eigenvalue $p$. Suppose that the pointer is in the state
\begin{math}
\ket{\phi_x}=\psi_x(\ket{0}+\ket{1}e^{i\theta p})
\end{math}
after the measurement of $\hatX$ on the system.
The two-state tomography can successfully extract the
relative phase factor $e^{i\theta p}$, and hence $p$. If $p$ is non-degenerate,
the eigenvalue itself uniquely identifies $\ket\psi$ as its eigenstate.
However, it is impossible if $p$ is degenerate.
Fortunately, this special case can be discerned experimentally
because $\varphi_x$ is independent of $x$, and
$P_{x,0}=P_{x,1}$ for all $x$.

The first two cases can be avoided simply by properly choosing either the
coupling operator $\hatP$ or the computational basis $\ket{x}$.

\subsection*{Iterative ML algorithm}

Here, we detail the maximization of the likelihood function over the entire
Hilbert space. Because of the normalization constraint, it is more convenient to
maximize
\begin{equation}
\label{TomographyPaper::eq:14}
\log\varL[\psi] - \lambda\sum_x|\psi_x|^2 \,,
\end{equation}
where $\lambda$ is the Lagrange multiplier.
Suppose that the system was initially in a definite state $\ket{y}$ and went through the unitary interaction $\hatU_\mathrm{int}$ with the pointer.
Let $\ket{\phi_{xy}}$ be the pointer state
upon the measurement outcome $x$ on the system.
Explicitly, it can be expressed as
\begin{equation}
\label{TomographyPaper::eq:15}
\ket{\phi_{xy}} := \ket{0}\delta_{xy} + \ket{1}V_{xy} \,.
\end{equation}
The pointer state $\ket{\phi_x}$ resulting from the general initial state
$\ket\psi$ of the system is related to $\ket{\phi_{xy}}$ by
\begin{math}
\ket{\phi_x} = \sum_y\ket{\phi_{xy}}\psi_y.
\end{math}
In terms of $\ket{\phi_{xy}}$, the joint probability can be expressed as
\begin{equation}
P_{x,m}
= \braket{\phi_x|\hat\Pi_m|\phi_x}
= \sum_{yz}\braket{m|\phi_{xy}}^*\psi_y^*\braket{m|\phi_{xz}}\psi_z
\end{equation}
For later use, it should be noted that its derivative with respect to $\psi_x$ has the form
\begin{equation}
\frac{\partial P_{x,m}}{\partial\psi_y^*}
= \sum_{z}\braket{m|\phi_{xy}}^*\braket{m|\phi_{xz}}\psi_z
\end{equation}
Then, the extremal equation for the maximization
problem~\eqref{TomographyPaper::eq:14} is given by
\begin{equation}
\label{TomographyPaper::eq:16}
\frac{\log\varL}{\partial\psi_y^*}
= \sum_{x}\sum_{m\in\calM}\frac{F_{x,m}}{P_{x,m}}
\frac{\partial P_{x,m}}{\partial\psi_y^*}
= \sum_{xz}\sum_{m\in\calM}
\braket{\phi_{xy}|m}
\frac{F_{x,m}}{P_{x,m}}
\braket{m|\phi_{xz}}\psi_z
= \psi_y \,.
\end{equation}
We define an $x$-dependent operator $\hatR_x$ on the pointer by
\begin{equation}
\hatR_x
:= \sum_{m\in\calM} \frac{F_{x,m}}{P_{x,m}}
\ket{m}\bra{m}
= \sum_{m\in\calM} \frac{F_{x,m}}{P_{x,m}}\hat\Pi_m .
\end{equation}
Then, the extremal equation~\eqref{TomographyPaper::eq:16} is
\begin{equation}
\sum_{xz}\bra{\phi_{xy}}\hatR_x\ket{\phi_{xz}}\psi_z = \psi_y \,.
\end{equation}
Putting~\eqref{TomographyPaper::eq:15} into the above equation, we obtain
\begin{equation}
\sum_{xz}\left(\delta_{yx}\bra{0}+V_{yx}^\dag\bra{1}\right)
\hatR_x
\left(\ket{0}\delta_{xz}+\ket{1}V_{xz}\right)
\psi_z = \psi_y \,,
\end{equation}
which is identical to the matrix equation~\eqref{TomographyPaper::eq:8}.

\subsection*{Matrix product states and operators}

Consider a system of $n$ particles, each of which has Hilbert space dimension
$d$.  We denote the computational basis state $\ket{x}$ for $x=0,1,\cdots,d^n-1$ as
\begin{math}
\ket{x} = \ket{x_1}\otimes\ket{x_2}\otimes\cdots\ket{x_n},
\end{math}
where $x_j$ are the base $d$ digits in $x$,
\begin{math}
x = x_1 + x_2 d + \cdots + x_n d^{n-1}.
\end{math}

An open boundary matrix product state (MPS) \cite{Perez-Garcia07a,Schollwock11a} is represented by
\begin{equation}
\ket\eta = \sum_{x}\ket{x_1}\otimes\ket{x_2}\otimes\cdots\otimes\ket{x_n}
A_1^{x_1}A_2^{x_2}\cdots A_n^{x_n} \,,
\end{equation}
where $A_j^{x_j}$ are the $D_j\times D_{j+1}$ complex matrices, depending on the
local state $x_j$, and $D_1=D_{N+1}=1$.
Similarly, an open boundary matrix product operator (MPO) takes the form
\begin{equation}
\hatO = \sum_{\mu_1=1}^{d^2}\sum_{\mu_2}\cdots\sum_{\mu_n}
\hat\tau_1^{\mu_1}\otimes
\hat\tau_2^{\mu_2}\otimes\cdots\otimes
\hat\tau_n^{\mu_n}\,
B_1^{\mu_1}
B_2^{\mu_1}\cdots
B_n^{\mu_1} \,,
\end{equation}
where $\hat\tau_j^{\mu_j}$ are the basis operators of the Hilbert space of all
linear operators acting on particle $j$; and $B_j^{\mu_j}$ are
$D_j'\otimes D_{j+1}'$ complex matrices ($D_1'=D_{n+1}'=1$).

One can observe that the conditional operator
\begin{math}
\sum_x\ket{x}\bra{x}\otimes\hatR_x[\psi^{(k)}]
\end{math}
is an MPO with a finite bond dimension
provided that the state $\ket{\psi^{(k)}}$ is an MPS with a finite bond
dimension. Because an MPS has finite correlations, the probabilities
$P_{x_1\dots x_n,m}$ are factorized as they are statistically independent of
the uncorrelated parts;\cite{Perez-Garcia07a,Schollwock11a} we recall the base-$d$ digits representation of $x$.
This is also the case for the experimental observed frequencies
$F_{x_1\dots x_n,m}$. Therefore, the conditional operator is an MPO with
a finite bond dimension.

\section*{Acknowledgements}

This work was supported by the National Research Foundation of Korea (Grant
Nos. 2017R1E1A1A03070681 and 2022M3H3A1063074) and by the Ministry of
Education of Korea through the BK21 program.

\bibliography{TomographyPaper}

\begin{thebibliography}{10}
\bibitem{endnote:2}
\bibinfo{note}{Physically, one may find the form $\hatU_\mathrm{int} =
  \exp\left(-i\theta\hatP\otimes\hat\sigma^z\right)$ more common. It only
  modifies the post-processing procedure slightly but not the essential part of
  the protocol.}
\end{thebibliography}

\end{document}